# Regulatory conformational changes of the ε subunit in single FRET-labeled F$_o$F$_1$-ATP synthase

Thomas M. Duncan[a], Monika G. Düser[b], Thomas Heitkamp[c], Duncan G. G. McMillan[c] and Michael Börsch[c,*]

[a] Department of Biochemistry & Molecular Biology, SUNY Upstate Medical University, Syracuse, NY, USA;
[b] 3$^{rd}$ Institute of Physics, Stuttgart University, Stuttgart, Germany;
[c] Single-Molecule Microscopy Group, Jena University Hospital, Friedrich Schiller University, Jena, Germany

## ABSTRACT

Subunit ε is an intrinsic regulator of the bacterial F$_o$F$_1$-ATP synthase, the ubiquitous membrane-embedded enzyme that utilizes a proton motive force in most organisms to synthesize adenosine triphosphate (ATP). The C-terminal domain of ε can extend into the central cavity formed by the α and β subunits, as revealed by the recent X-ray structure of the F$_1$ portion of the *Escherichia coli* enzyme. This insertion blocks the rotation of the central γ subunit and, thereby, prevents wasteful ATP hydrolysis. Here we aim to develop an experimental system that can reveal conditions under which ε inhibits the holoenzyme F$_o$F$_1$-ATP synthase *in vitro*. Labeling the C-terminal domain of ε and the γ subunit specifically with two different fluorophores for single-molecule Förster resonance energy transfer (smFRET) allowed monitoring of the conformation of ε in the reconstituted enzyme in real time. New mutants were made for future three-color smFRET experiments to unravel the details of regulatory conformational changes in ε.

**Keywords:** F$_o$F$_1$-ATP synthase; ε subunit; conformational change; single-molecule FRET.

## 1 INTRODUCTION

To synthesize adenosine triphosphate (ATP) from adenosine diphosphate (ADP) and inorganic phosphate (P$_i$), the *Escherichia coli* enzyme F$_o$F$_1$-ATP synthase utilizes the electrochemical potential of protons. The bacterial enzyme can also work in reverse and can hydrolyze ATP to pump protons across the membrane[1]. Mechanical rotation of subunits couples proton translocation as the driving force in the F$_o$ portion with chemical synthesis or hydrolysis of ATP in the F$_1$ portion of the enzyme. This mechanism was first proposed by P. Boyer about 30 years ago (reviewed in [2]). It was demonstrated subsequently by a variety of biochemical[3-7] and spectroscopic[8] methods as well as single-molecule imaging[9-17] and single-molecule FRET experiments[18-28].

The crystal structure of the *Escherichia coli* F$_1$ portion was recently resolved at a resolution of 3.26 Å[29, 30]. F$_1$ consists of five different subunits with stoichiometry α$_3$β$_3$γδε. The pseudohexagonal arrangement of three pairs of subunits α and β, i.e. α$_3$β$_3$, forms the main body of F$_1$. Each β subunit provides an active nucleotide binding site. The corresponding nucleotide binding sites on the α subunits are catalytically inactive. Subunits α$_3$β$_3$ together with subunit δ at the top of F$_1$ comprise a non-rotating stator complex[31]. Subunits γ and ε form the central stalk that can rotate within α$_3$β$_3$ and connects to the membrane-embedded ring of 10 *c*-subunits of the F$_o$ portion (Fig. 1A). The shape of this large enzyme is shown in Fig. 1A as an image reconstruction from electron micrographs[32]. In the F$_o$ portion, the *a* subunit provides two half-channels for proton translocation across the membrane. Two *b* subunits of F$_o$ connect the membrane part as a peripheral stator stalk to the top of F$_1$ (Fig. 1A). Thus, the holoenzyme $ab_2c_{10}α_3β_3γδε$ can transfer the energy of the transmembrane electrochemical potential of protons *via* rotational movements of $c_{10}$-ε-γ to the distant nucleotide binding sites in α$_3$β$_3$ where ATP is synthesized.

...................................................................................

* Email: michael.boersch@med.uni-jena.de; http://www.m-boersch.org

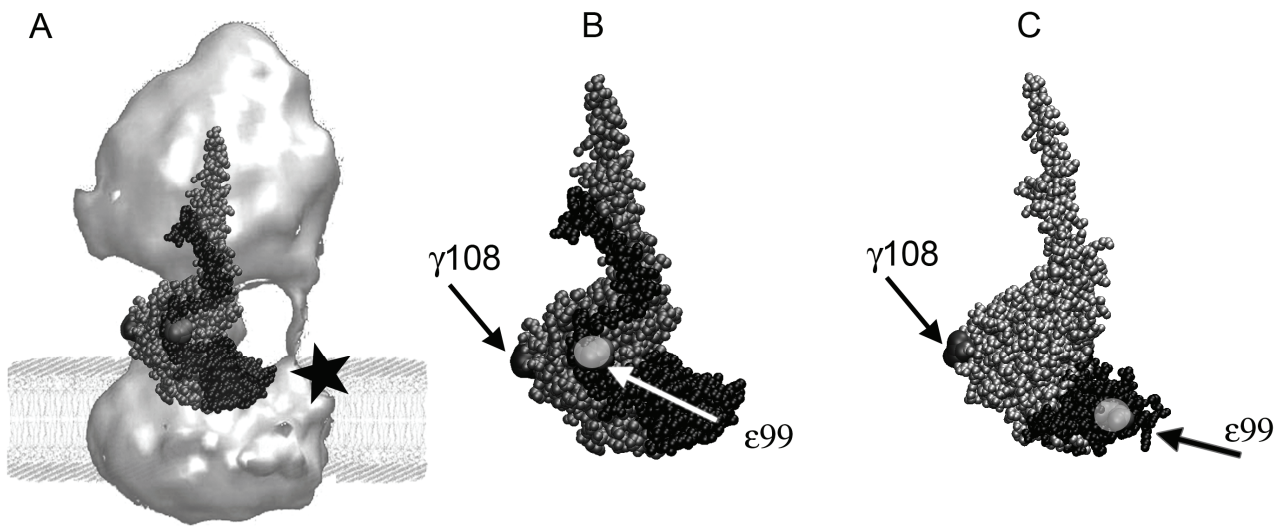

**Figure 1.** (**A**) Structure of *E. coli* $F_oF_1$-ATP synthase (surface rendering from electron micrographs[32]) in a model lipid bilayer. Overlaid here in ball representation are the subunits that comprise the central rotary stalk: γ (grey) and ε (black) of the recent *E. coli* $F_1$ X-ray structure[29]. Subunit *a* with a C-terminal cysteine mutation (black star) can be labeled for future 3-color FRET experiments. (**B**) Subunits γ (in grey) and ε (in black, in the 'up' configuration [29]), showing the positions of two cysteine mutations γ108 and ε99 (highlighted in light grey) for single-molecule FRET. (**C**) Partial structure of γ (in silver) and ε (in black) from $F_1$ with ε's C-terminal helices in the 'down'-configuration[33]. Compared to (**B**), note the distinct distance between cysteines γ108 and ε99 for single-molecule FRET.

The bacterial $F_oF_1$-ATP synthase is thought to be regulated by conformational changes in subunit ε, a 15 kDa subunit of the $F_1$ rotor[34], to control and prevent wasteful ATP hydrolysis *in vivo*. An *E. coli* $F_oF_1$-ATP synthase with a deleted C-terminal domain (CTD) of subunit ε showed not only a higher ATP hydrolysis (ATPase) activity compared to the wild type but also higher ATP synthesis activity[35]. With the CTD of ε in an 'extended' configuration (Fig. 1B), the enzyme is in an intrinsically inactive state with the γ-ε rotor stalled at a fixed angle[29, 36, 37]. However, ε's CTD can also form a hairpin-folded state[33, 38] with the C-terminal helices in a 'down'-configuration (Fig. 1C), and the membrane-bound enzyme can catalyze ATP synthesis and hydrolysis with ε trapped in that state[39]. Therefore, a large rearrangement of the C-terminal helices of ε is thought to be a mechanical switch that controls the enzymatic activities of both $F_1$-ATPase as well as $F_oF_1$-ATP synthase.

We developed a single-molecule FRET approach to monitor conformational changes of ε's CTD in purified *E. coli* $F_oF_1$-ATP synthase reconstituted in liposomes. Two specifically attached fluorophores were used for smFRET as an internal distance ruler. Based on the *E. coli* $F_1$ X-ray structure[29], we chose residue ε99 on the first C-terminal α-helix of ε, which does not insert into a β-γ cleft in the 'up'-conformation. The second marker position is γ108, yielding distances of about 3 nm or 6 nm to ε99 in the 'up' (Fig. 1B) or 'down' (Fig. 1C) conformations, respectively. These labeling positions were also chosen to avoid perturbing any interactions of ε's CTD (either conformation) with the ε N-terminal domain (NTD) or with the other subunits γ and *c*. We recorded the dynamics of ε's CTD for several hundred milliseconds using freely diffusing proteoliposomes with single FRET-labeled $F_oF_1$-ATP synthases.

# 2 EXPERIMENTAL PROCEDURES

## 2.1 Preparation of the $F_1$ portion and the isolated δ and ε subunits

The soluble $F_1$ portion ($F_1$) of *E. coli* ATP synthase with an N-terminal 6xHistidine tag ($H_6$) on the β subunits and a specific cysteine substitution in the γ subunit (γK108C) was isolated from membranes and depleted of subunits δ and ε. $H_6$-tagged ε subunit was overexpressed separately and contained a unique cysteine mutation εR99C for labeling. The δ subunit was overexpressed separately and purified to complete the $F_1$ portion. Details of the plasmid constructs used, protein expression and purification procedures are given below.

*Plasmid constructs*. Plasmid pJW1[40] was the basis for the construct used to overexpress $F_oF_1$ for purification of $F_1$. The $H_6$-tag at the N-terminus of β was engineered previously[41]. All native Cys of β and γ subunits were changed to Ala by replacing appropriate restriction fragments of pJW1 with the corresponding regions of a similar plasmid encoding Cys-free $F_OF_1$[42]. The γK108C mutation (AAA > TGC) was made by site-directed mutagenesis (QuikChange II, Agilent Technologies). Plasmid $pH_6ε$[43] was mutated to express $H_6$-tagged ε with εR99C (<u>C</u>GT > <u>T</u>GT) and contained no other cysteines. The plasmid $pH_6ε$ comprises the amino terminal ($His_6$) and a rTEV protease site sequence, MSYYHHHHHH-DYDIPTTENLYFQGA, preceding the *atpC* open reading frame[43]. Plasmid pJC1[44] was used for separate expression of the wild-type δ subunit. For all tag sites and mutated regions, DNA sequencing was done to confirm the presence of the expected tag/mutant sequence and the absence of any undesired sequence changes.

*Expression of $F_oF_1$ and purification of $F_1(H_6β/γK108C)$ depleted of δ and ε subunits.* For overexpression of the engineered ATP synthase, pJW1($H_6β$/γK108C) was transformed into an *E. coli* strain that lacks a chromosomal *atp* operon (LE392Δ(*atpI-C*) [45]). Cells were grown in up to 10 L of defined medium[40] supplemented with 50 mg Met/L, and everted membranes were prepared[40]. Soluble $F_1(H_6β/γK108C)$ was released from membranes and purified as described[29]. The ε subunit was depleted from $F_1$ with an anti-ε immunoaffinity column[46] as described[36]. To remove residual impurities and deplete most δ subunit, $F_1$ was subjected to immobilized metal affinity chromatography at 4°C. $F_1(H_6β/γK108C)$ was diluted in Talon buffer (50 mM Tris-HCl, 40 mM 6-amino n-hexanoic acid, 10%(v/v) glycerol, 5 mM β-mercaptoethanol (βME), 1 mM ATP; pH adjusted to 7.2 at 22°C) and bound to a 10 ml column of Talon resin (Clontech). The column was washed at 2 ml/min with Talon buffer and different additions as follows: 2 volumes of buffer alone, 4 volumes plus 0.2%(w/v) lauryldimethylamineoxide (LDAO) to deplete most of subunit δ, 2 volumes of buffer alone to remove LDAO, and 3 volumes plus 0.1 M imidazole to elute $F_1(H_6β/γK108C, –δε)$. The final $F_1(–δε)$ sample was dialyzed against Talon buffer (with 1 mM DTT instead of βME) to remove imidazole and residual LDAO, then concentrated to >5 mg/ml by ultrafiltration (Vivaspin, 50 kDa MWC) and stored at -80°C.

*Expresssion and purification of $H_6$-tagged εR99C.* $His_6$-tagged ε subunit ($H_6εR99C$) was expressed in *E. coli* strain BL21Star(DE3) (Life Technologies). Cell growth and induction by IPTG were done as described[43]. The $H_6εR99C$ subunit was purified essentially as described for $H_6ε$[36] but buffers for Talon chromatography contained 1 mM βME, whereas buffers for gel filtration and storage contained 1 mM DTT.

*Expresssion and purification of subunit δ.* Plasmid pJC1 was transformed into strain LE392Δ(*atpI-C*). Cells were grown at 30°C in 2 Liters of LB broth plus 1 mM $MgSO_4$, 1%(v/v) glycerol, and ampicillin (0.1 mg/ml). Expression of δ was induced with IPTG and isolation of δ was done as described[44], through the step of anion exchange chromatography (BioRad, Macro-Prep DEAE resin, 10 ml column). This partially pure δ was concentrated and exchanged into 'TED' buffer (20 mM Tris-HCl, 1mM EDTA, 1 mM DTT, pH 8.0) with 0.2 M $Na_2SO_4$ and subjected to hydrophobic interaction chromatography at 4°C. The δ sample was applied to a 10 ml column of Macro-Prep t-butyl resin (BioRad), and a flow rate of 1 ml/min was used. The column was washed with 3 volumes of TED+0.2 M $Na_2SO_4$ and δ (>95% pure) eluted immediately. Importantly, this eliminated residual amounts of a ~14 kDa proteolytic fragment of δ that is often present[44]. For the final δ sample, ultrafiltration (Vivaspin, 10K MWCO) was used to reduce $Na_2SO_4$ to ~10 mM and concentrate δ to >15 mg/ml for storage at -80°C.

## 2.2 Labeling of $F_1(H_6β/γK108C)$ with Atto488-maleimide and $H_6εR99C$ with Atto647N-maleimide.

The γ subunit of $F_1$ with the mutation K108C was labeled with Atto488-maleimide as described[47, 48]. Fluorescence labeling of 13 μM $F_1$ resulted in a labeling ratio of 0.55 for the γ subunit according to quantitative SDS-PAGE analysis using a Typhoon scanner. In addition, cysteines of the residual δ subunit were partly labeled with Atto488. $F_1$-Atto488 was flash-frozen in liquid $N_2$ and stored at -80° C in MTKE7 buffer (20 mM MOPS-Tris, 50 mM KCl, 0.1 mM EDTA, pH 7.0) with 10% glycerol and 1 mM ATP. $His_6$-tagged ε subunit with the mutation R99C was labeled with Atto647N-maleimide. 50.1 µM ε contained 17.4 µM Atto647N yielding a labeling ratio of 0.30. ε-Atto647N was flash-frozen in liquid $N_2$ and stored in MTKE7 buffer with 10% glycerol at -80° C.

## 2.3 Preparation of *E. coli* $F_oF_1$ and $F_o$-proteoliposomes

*E. coli strain and growth conditions.* Plasmid constructs were based on plasmid pRA100 which carries the *atp* operon (without *atp*I)[49]. For expression of these *atp* genes, the strain RA1 (F- *thi rpsL gal* Δ(*cyoABCDE*)456::KAN Δ(*atpB-atpC*) *ilv*:Tn10)[50] was used lacking a functional $F_oF_1$-ATP synthase. Cells were grown in a modified complex medium (0.5 g/l yeast extract, 1 g/l tryptone, 17 mM NaCl, 10 mM glucose, 107 mM $KH_2PO_4$, 71 mM KOH, 15 mM $(NH_4)_2SO_4$, 4 µM uracil, 50 µM $H_3BO_3$, 1 µM $CoCl_2$, 1µM $MnCl_2$, 2 µM $ZnCl_2$, 10 µM $CaCl_2$, 3 µM $FeCl_2$, 0.5 mM $MgSO_4$, 0.5 mM arginine, 0.5 mM isoleucine, 0.7 mM valine, 4 µM thiamine and 0.4 µM 2,3-dihydroxybenzoic acid) in a 10 L fermenter (FerMac 320, Electrolab, UK). Growth conditions were as follows: 37°C, 800 rpm stirring, aerated with 4.5 L/min, harvested at the middle-late logarithmic phase ($OD_{600nm}$ = 0.8). Cells were collected by centrifugation in a Sorvall Evolution RC centrifuge (Thermo Fisher Scientific, USA) at 10,000 × g and 4° C for 5 min, washed with 50 mM Tris-HCl pH 8.0, and collected as described. Cells were then flash-frozen in liquid nitrogen and stored at -80°C.

*Purification of E. coli $F_oF_1$.* $F_oF_1$ was purified by a combination of two published methods. Membranes were isolated and washed using a modified protocol[51] followed by a modified solubilization and further purification as described[52]. Briefly, approximately 30 g frozen cells were thawed in a water bath and resuspended in 200 ml buffer A containing 50 mM MOPS-KOH pH 7.0, 175 mM KCl, 10 mM $MgCl_2$, 0.2 mM EGTA, 0.2 mM DTT and 0.1 mM PMSF. Unless otherwise noted, all subsequent steps were done at 4°C. One spatula point of DNase I and RNase A were added directly before cell lysis by two passages through a precooled PandaPlus 2000 cell homogenizer (GEA Niro Soavi, Italy) at 1000 bar. Between the two passages, the cells were kept on ice. After removal of cell debris by two centrifugations at 25,000 × g for 20 min in a Sorvall Evolution RC centrifuge (Thermo Fisher Scientific, USA), membranes were collected by centrifugation at 300,000 × g for 2 h in an Optima XP ultracentrifuge (Beckman Coulter, USA) using a type 70 Ti rotor. The membrane pellet was resuspended in 150 ml buffer B containing 5 mM Tris-HCl pH 8.0, 5 mM $MgCl_2$, 0.2 mM EGTA, 0.2 mM DTT, 6 mM p-amino-benzamidine (PAB), 10%(v/v) glycerol and 0.1 mM PMSF using a soft paintbrush and centrifuged for 1.5 h at 300,000 × g. This washing step was repeated using 100 ml buffer B. The pellet was homogenized in 20 mM MES/Tricine-KOH pH 7.0, 5 mM $MgCl_2$, 2 mM DTT and 0.001% (w/v) PMSF and adjusted to 10 ml buffer per g membrane protein. $F_oF_1$ was solubilized by dropwise addition of dodecylmaltoside (DDM; Glycon, Germany; 15% (w/v) stock solution) to 1.75% (w/v) final, then incubating for 2 h on ice with gentle stirring. Unsolubilized membranes were then pelleted by centrifugation at 300,000 × g for 1.5 h. Ammonium sulfate was added in two steps to the supernatant. Impurities were precipitated first with 45% (w/v) saturated $(NH_4)_2SO_4$ and separated by centrifugation (35,000 x g, 15 min). Then, $F_oF_1$ was precipitated with 65% (w/v) saturated $(NH_4)_2SO_4$ and pelleted in the same way.

The protein pellet was dissolved in 2.5 ml of size exclusion buffer (SEC-buffer) containing 40 mM MOPS-KOH pH 7.5, 80 mM KCL, 4 mM $MgCl_2$, 2 mM DTT, 2% (w/v) sucrose, 10% (v/v) glycerol, 1% (w/v) DDM and 0.001% (w/v) PMSF. To separate the protein from residual $(NH_4)_2SO_4$, lipids and nucleotides, the solution was applied to a self-packed XK16/100 Sephacryl S300 size exclusion column connected to an Äkta PrimePlus FPLC (GE-Healthcare, USA) and equilibrated with SEC-buffer containing 0.1% (w/v) DDM. The protein was eluted at a flow rate of 0.6 ml/min and fractions of 4 ml were collected. The peak fractions were pooled to 8 ml fractions and were loaded separately on a Poros HQ 20 (4.6 x 100 mm) ion exchange column (Applied Biosystems, USA) connected to a modular FPLC system (LCC500 controller, P500 pumps, Frac100 fractionator, UVM II UV monitor) (Pharmacia, Sweden) and equilibrated with the SEC-buffer. After washing the column with 5 column volumes of SEC-buffer, $F_oF_1$ was eluted by a KCL gradient over 20 column volumes up to 0.75 M KCl and collected in 1 ml fractions. The protein of the pooled main peak fractions was precipitated with 65% (v/v) saturated $(NH_4)_2SO_4$ and pelleted by centrifugation (35,000 × g, 15 min). The protein pellet was then resolved in 0.5 ml SEC-buffer. To separate the $F_oF_1$ from residual salt and minor impurities, the sample was applied to a second size exclusion step using a self-packed XK16/100 Sephacryl S400 column (GE-Healthcare, USA) and the buffers and specifications listed above. The peak fractions were directly tested for ATPase activity, pooled and shock-frozen in liquid nitrogen in 500 µl cryo straws and stored at -80°C.

*Reconstitution.* $F_oF_1$ was reconstituted into preformed liposomes as described[53] at a concentration of 20 nM $F_oF_1$. Given a lipid concentration of 8 g/l and a mean liposome diameter of 120±10 nm, a ratio of four liposomes per $F_oF_1$ ensured proteoliposomes with only a single enzyme[18]. Briefly, 2.5 µl of a 1 M $MgCl_2$ solution, 500 µl of preformed liposomes[54] and 10.6 µg of purified protein were diluted with liposome buffer (20 mM Tricine-NaOH pH 8.0, 20 mM succinate, 80 mM NaCl and 0.6 mM KCl) to a volume of 920 µl. Immediately, 80 µl of a 10% (v/v) Triton X-100 solution were added under vigorous stirring to destabilize the liposomes. After 1 h incubation under slow shaking (2 rpm, 45° angle) 520 mg of BioBeads SM-2 (Biorad, USA), pretreated according to[55], were added to remove the detergent. After an additional hour of shaking, the proteoliposomes were separated from the BioBeads and immediately used for stripping off the $F_1$ portion of the enzyme.

*Stripping of $F_1$.* $F_o$-liposomes were prepared by stripping the $F_1$ portion (modified from[56]). 1 ml $F_oF_1$-liposomes were diluted with 24 ml of stripping buffer (1 mM Tricine–NaOH pH 8.0, 1 mM DTT, 0.5 mM EDTA, and 4% (v/v) glycerol) in a type 70 Ti centrifuge tube (Beckman Coulter, USA), incubated for 1 h at room temperature under slow rotation (5 rpm), and centrifuged at 300,000 × g for 1,5 h at room temperature. The pellet was resuspended twice in 25 ml of stripping buffer followed by incubation and centrifugation under the same conditions. Finally, the stripped $F_o$-liposomes were resuspended in stripping buffer with 10% glycerol to yield an $F_o$ concentration of 125-150 nM, flash-frozen in 10 µl aliquots in liquid nitrogen and stored at -80°C.

*Other methods.* Protein concentrations were determined either by UV absorption or by the amido black method[57]. SDS-polyacrylamide gels were made according to[58] with a 12% separating gel and either stained by silver[59] or coomassie R-250[60].

## 2.4 Preparation of proteoliposomes with a single $F_oF_1$-γ108-atto488/ε99-atto647N

*Rebinding of FRET-labeled $F_1$ to $F_o$-proteoliposomes.* Atto488-labeled $F_1$(-δε) (see above) was reassembled with Atto647N-labeled ε in a first step and subsequently rebound to Fo-liposomes in two steps. First, 4 µM Atto647N-labeled ε was bound to 3 µM of Atto488-labeled $F_1$(–δε) by incubation in liposome buffer for 30 min at room temperature yielding FRET-labeled $F_1$(–δ). A fivefold excess of the purified δ subunit was added to complement $F_1$. FRET-labeled $F_1$ in solution was incubated with the $F_o$-liposome suspension at a molar excess of three $F_1$ per $F_o$ in the presence of 2.5 mM $MgCl_2$ and 50 mM NaCl, first for 45 min at 37°C and then for 90 min at 0°C. Excess, unbound $F_1$ and δ were removed by three ultracentrifugations (90 min, 300,000 × g, 4°C), each time resuspending the pellet in buffer (20 mM Tricine-NaOH (pH 8.0), 20 mM succinic acid, 50 mM NaCl, 0.6 mM KCl, 2.5 mM $MgCl_2$ and 4% (v/v) glycerol). The final concentration of $F_oF_1$ in proteoliposomes was adjusted to ~100 nM. Proteoliposomes were either used directly for smFRET measurements, or adjusted to 10% (v/v) glycerol, flash-frozen as 10 µl aliquots in liquid nitrogen and stored at -80°C.

## 2.5 Confocal single-molecule FRET microscope

*Custom-designed confocal microscope for smFRET.* The setup has been described previously[61, 62]. Briefly, a solid-state continuous-wave laser (Cobolt Calypso, 50 mW) was used for excitation with 491 nm. The laser beam was deflected by the two crystals of a pair of acousto-optical beam deflectors (AOBD 46080-3-LTD, NEOS technologies, Gooch & Housego). The diffracted first order beam was selected by an aperture. The beam diameter was diminished by two lenses ($f_1$= 200 mm, $f_2$= 100 mm) before entering the back aperture of an oil immersion microscope objective (PlanApo 100x, N.A. 1.35, Olympus). A dichroic beam splitter (488 RDC, AHF Tübingen, Germany) rejected scattered laser light from fluorescence photons in the detection pathway. An achromatic lens (200 mm) imaged the confocal excitation volume onto the center of the pinhole (150 µm) which was mapped on to the detection area of two single photon-counting avalanche photodiodes (SPAD, SPCM-AQR-14, Perkin-Elmer) by additional lenses (50 mm). Fluorescence was separated into two channels for FRET measurements by a beam splitter 575DCXR (AHF Tübingen). FRET donor fluorescence was detected after an additional band pass filter ET535/70M (AHF Tübingen), and FRET acceptor fluorescence was detected after a long pass filter HQ 595 LP (AHF Tübingen). Arrival times of photons were recorded with TCSPC electronics (16 channel photon correlator DPC230, Becker&Hickl, Berlin, Germany). However, the picosecond time resolution of the TCSPC card was not required for the measurements of binned fluorescence intensity only. Data analysis was performed with the software 'Burst Analyzer' (version 2.0, Becker&Hickl, Berlin, Germany).

Single-molecule FRET experiments were carried out in liposome buffer (20 mM succinic acid, 20 mM tricine, 80 mM NaCl, 0.6 mM KCl, 2.5 mM $MgCl_2$, adjusted pH to 8.0 with NaOH). Proteoliposome aliquots were used within 24 h after thawing.

# 3 RESULTS

To study the conformations of the C-terminal helices of the ε subunit in $F_oF_1$-ATP synthase by single-molecule FRET, we first attached the two FRET fluorophores specifically to the $F_1$ portion and then reassembled the holoenzyme by binding $F_1$ to $F_o$ in liposomes in the presence of subunit δ. Cysteine 108C of the γ subunit in (ε- and δ-depleted) $F_1$ was labeled with Atto488 with a labeling efficiency of 55%. The ε subunit was purified separately, and cysteine residue ε99C was labeled with Atto647N-maleimide as FRET acceptor with 30% efficiency as described above (and elsewhere[47]). Mixing $F_1$ (3 μM) with ε (4 μM) yielded FRET-labeled $F_1$, due to ε's high binding affinity ($K_D$~ 0.3 nM[36]). In the presence of a fivefold excess of purified δ, labeled $F_1$ was reassembled with non-labeled $F_o$ in liposomes to yield the holoenzyme $F_oF_1$-ATP synthase as described above and according to published procedures[18].

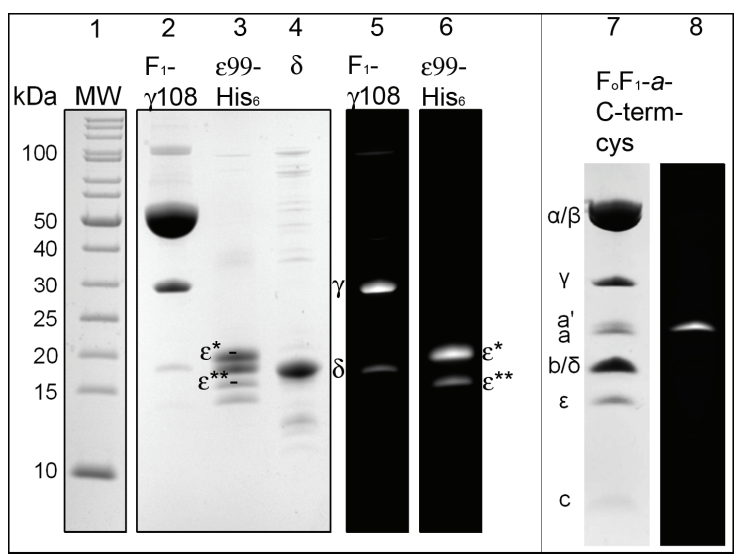

**Figure 2. Lanes 1 to 6**: SDS-PAGE of labeled $F_1$-γ108-atto488 and $His_6$ε99-atto647N (12% PA gel according to Schaegger and von Jagow[58], with Coomassie blue staining of $F_1$-ATPase subunits in lanes 2 to 4, and fluorescence images in lanes 5 and 6. Lane 1 is the molecular weight standard. Lane 2 shows $F_1$-γ108-atto488. Atto488-maleimide labeling resulted in an additional labeling of cysteines of residual δ subunit (lane 5). Lane 3, ε99-atto647N preparation showing two fluorescence-labeled products ε* and ε** (lane 6, for details see text). **Lanes 7 and 8**: SDS-PAGE of labeled $F_oF_1$-a-C-Term-cys. Lane 7 shows Coomassie blue-stained subunits of $F_oF_1$-ATP synthase, lane 8 specific Alexa488-maleimide labeling of the cys residue introduced to the C-terminus of subunit *a* (for details see text).

Figure 2 shows the different protein preparations as separated subunits on SDS-PAGE, either as fluorescence images after labeling or subsequent staining with Coomassie blue. $F_1$-(δ,ε) was mainly labeled on γ108 (lane 5), with minor labeling of residual δ that contained the two native cysteines. Labeling ε yielded two fluorescent products ε* and ε** (lane 6). Because ε contained a $H_6$-tag for purification and an additional TEV cleavage site, the bright ε* band likely represents the Atto647N-labeled, full-length tagged subunit, and ε** a small fraction of labeled ε that has been proteolyzed at 1 or more undetermined sites. Using SDS-PAGE according to Schaegger and von Jagow[58] showed separate bands for Atto647N-labeled and unlabeled ε subunit (bands below ε* or ε** in lane 3). This is likely due to the additional molecular weight (770 Da), additional positive charge and the hydrophobicity of the dye. On Laemmli-type gels[63] (not shown) only two prominent bands were seen, and both showed the Atto647N labeling. Similarly, the Alexa488-labeled *a* subunit with a cysteine added to the C-terminus (*a*-GAAACA) in $F_oF_1$-ATP synthase was separated as *a'* from unlabeled *a* in SDS-PAGE according to Schaegger and von Jagow (lanes 7, 8 in Fig. 2). We used $F_1$-(δ,ε) and ε without further purifications for generating the reconstituted FRET-labeled holoenzyme $F_oF_1$-ATP synthase.

Single-molecule FRET time trajectories were recorded with a custom-designed confocal microscope using 491 nm continuous-wave excitation (260 μW at the back aperture of the objective). The mean observation time of freely diffusing $F_oF_1$-ATP synthase in liposomes with ~120 nm diameter was 30 to 70 ms, compared to a diffusion time $t_D$~410 µs for rhodamine 110 in water (FCS data not shown). Maximum photon counts for single rhodamine 110 molecules in 1-ms-binned time traces were ~80 kHz, or 80 counts per ms, respectively. Accordingly the upper limit to assign a single FRET-labeled $F_oF_1$ ATP synthase in a liposome was set to 100 counts/ms for further analysis.

Fig. 3 shows six examples of photon bursts from single $F_oF_1$-ATP synthases in the absence of added nucleotides. Fluorescence intensities of FRET donor ($I_D$, dark grey traces) were corrected for 4 kHz background, FRET acceptor intensities ($I_A$, light grey traces) were corrected for 10 kHz background. FRET efficiencies were calculated as proximity factors $P=I_A/(I_D+I_A)$, without corrections for detections efficiencies or quantum yields of the fluorophores Atto488 and Atto647N, and are shown in the upper trace of each panel. Zooming in will show details of these screen shots from the 'Burst Analyzer' software including the recording time window of the photon burst. For clarity, proximity factors outside the marked bursts are masked, and the mean intensity-weighted P values for the marked bursts are plotted as black lines.

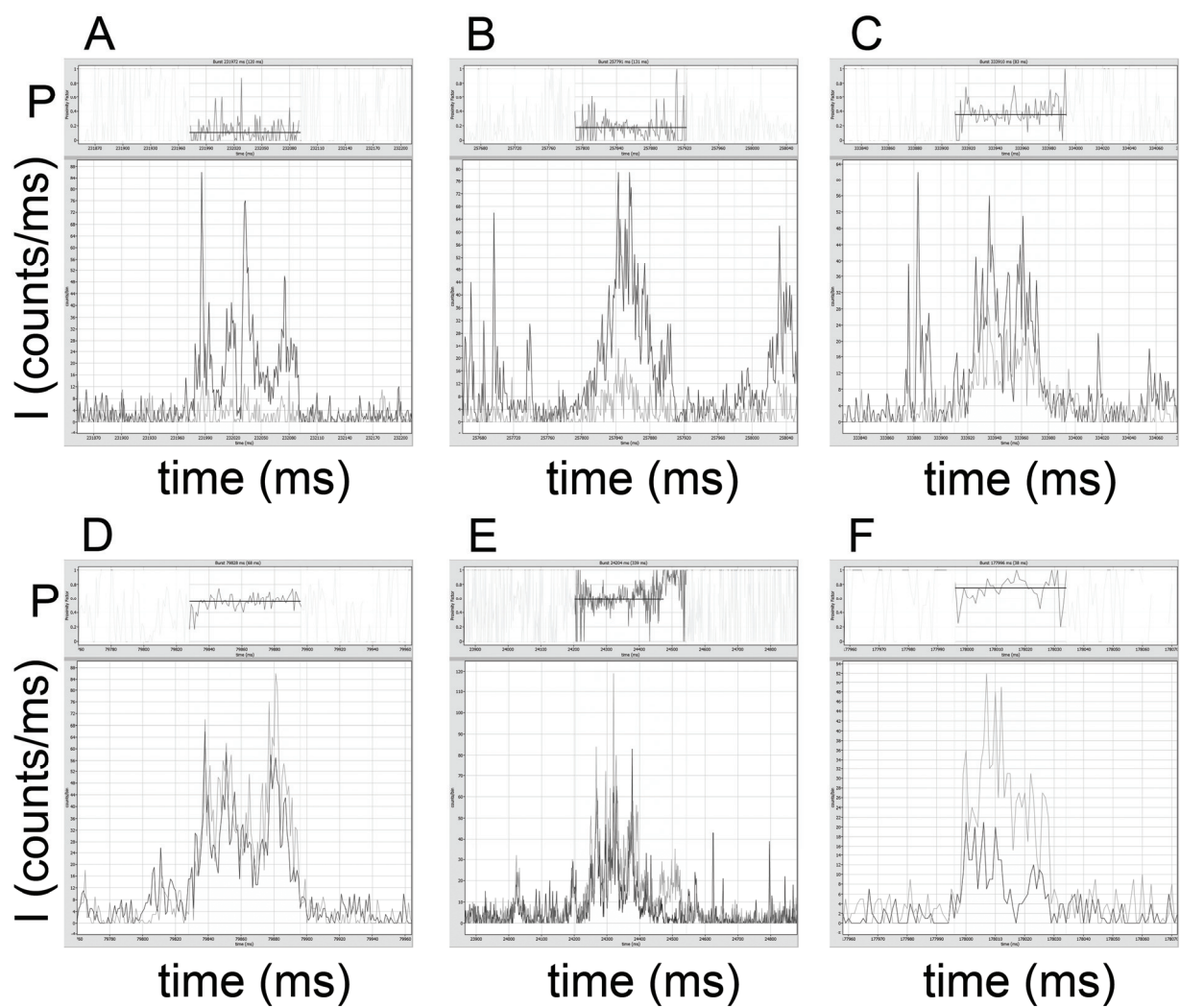

**Figure 3**. Photon bursts of FRET-labeled $F_oF_1$-ATP synthases in the absence of nucleotides. Atto488 intensities are shown in dark grey, Atto647N intensities in light grey. Intensities are given in counts per ms. Corresponding proximity factor P traces within the marked bursts are found in the upper part of the images, with mean intensity-weighted P values as black lines (see text for photon burst details).

A broad distribution of constant proximity factor values was found. In Fig. 3A, the observation time of the photon burst was 120 ms with P~0.12. In Fig. 3B the burst duration was 131 ms with P~0.17, and in Fig. 3C the burst was observed for 83 ms with P~0.35. Higher FRET efficiencies were found as well: in Fig. 3D with P~0.56 (68 ms duration), in Fig. 3E with P~0.59 (339 ms) and in Fig. 3F with P~0.74 (38 ms). For each example, standard deviation for the P value was in the range of ~0.1.

However, we also observed photon bursts with fluctuating, stepwise switching proximity factors and oligomers of FRET-labeled enzyme in liposomes as shown in Fig. 4. Some bursts were characterized by P ~0.9 to 1 with no detectable FRET donor signal (Fig. 4A; burst duration 202 ms), which we interpreted as likely artifacts.

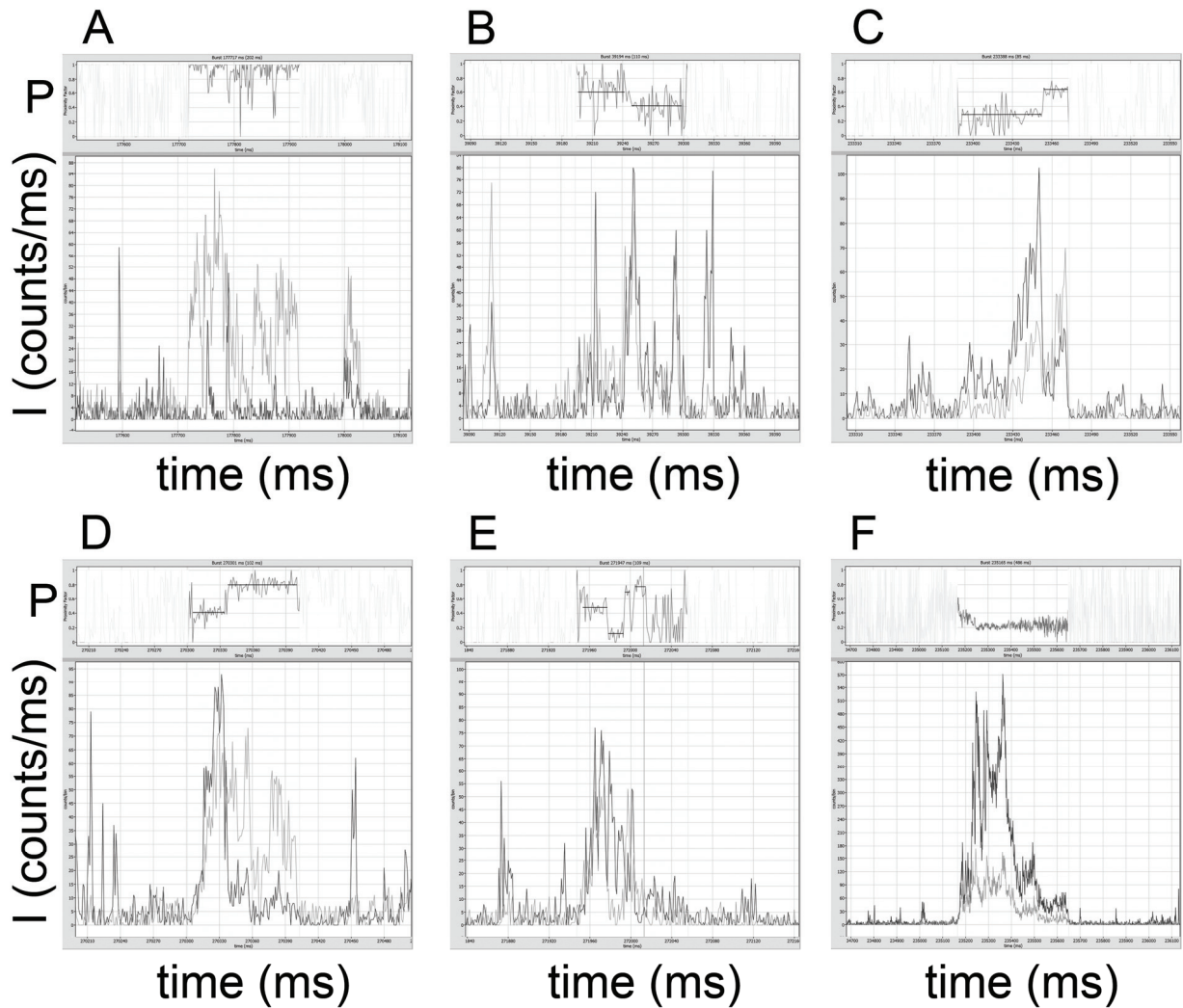

**Figure 4**. Photon bursts of FRET-labeled $F_oF_1$-ATP synthases in the absence of nucleotides. Atto488 intensities are shown in dark grey, Atto647N intensities in light grey. Intensities are given in counts per ms. Corresponding proximity factor P traces within the marked bursts are found in the upper part of the images, with mean intensity-weighted P values as black lines (see text for photon burst details).

In Fig. 4B, in a photon burst with 110 ms duration, stepwise FRET changes occurred from P~0.60 (for 49 ms) to P~0.41 (50 ms) with manual assignment of apparently constant FRET levels. The photon burst with 85 ms duration in Fig. 4C switched from P~0.28 (62 ms) to P~0.63 (18 ms), and another in Fig. 4D from P~0.41 (29 ms) to P~0.80 (61 ms) during an observation time of 102 ms. The 109 ms long photon burst in Fig. 4E showed multiple changes, from P~0.47 (23 ms) to P~0.13 (15 ms) to P~0.71 (4 ms) and to P~0.77 (10 ms) after an intermediary, brief low-FRET period. All photon bursts were characterized by photon count rates of less than 100 counts per ms for FRET donor plus FRET acceptor. Therefore, the few bursts with intensities exceeding 500 counts per ms, as shown in Fig. 4F, clearly originated from multiple FRET-labeled $F_oF_1$-ATP synthases in single liposomes. These multimers were rarely detected and were eliminated from further analysis.

Next we recorded single-molecule FRET of $F_oF_1$-ATP synthase in the presence of 1 mM MgATP. Only a few long-lasting photon bursts exhibited a single constant FRET level; most $F_oF_1$-ATP synthases showed transitions in FRET efficiencies as shown in Fig. 5.

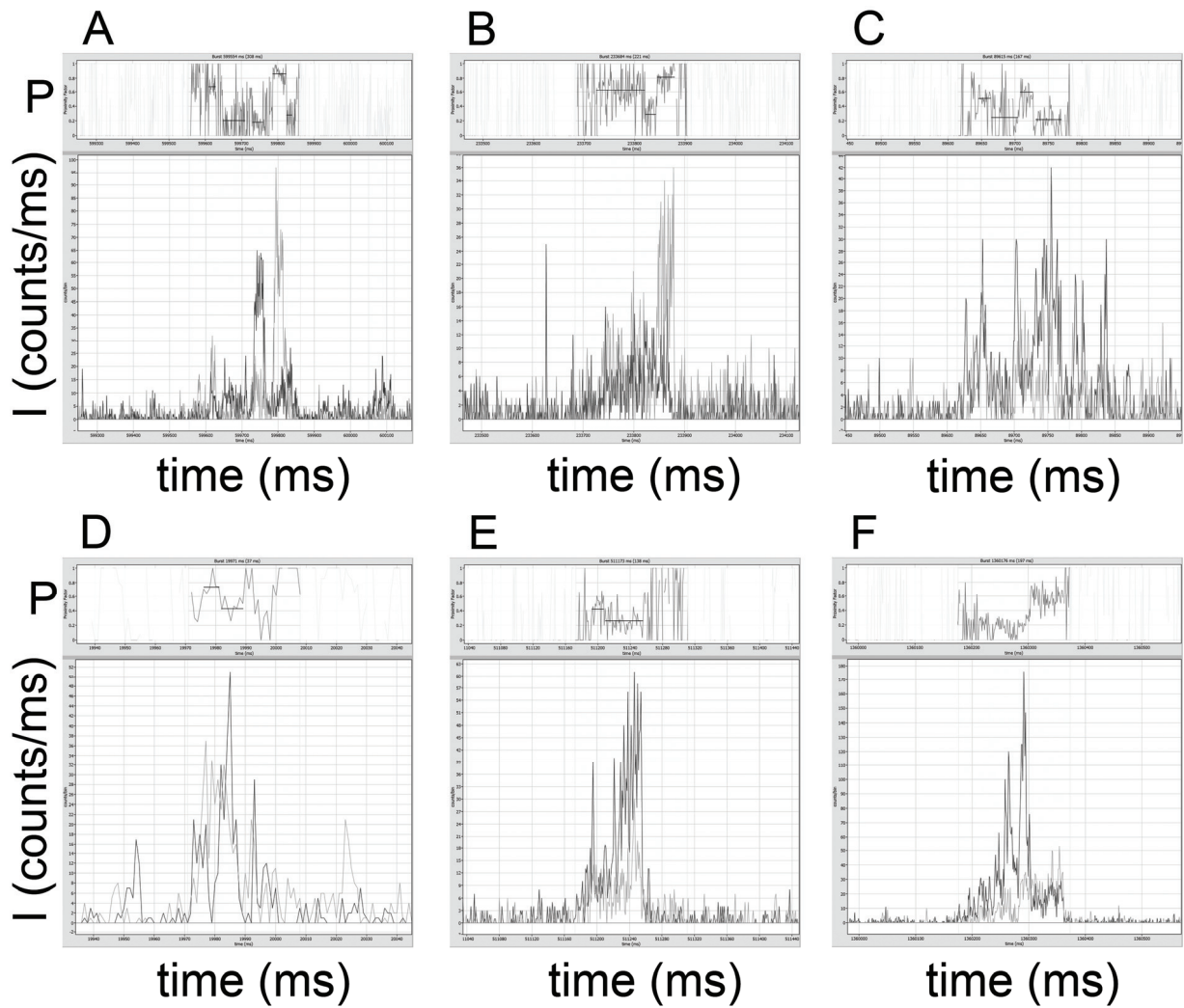

**Figure 5.** Photon bursts in the presence of 1 mM MgATP with fluctuating proximity factors P. Atto488 intensities are shown in dark grey, Atto647N intensities in light grey. Intensities are given in counts per ms. Corresponding proximity factor P traces within the marked bursts are found in the upper part of the images, with mean intensity-weighted P values as black lines (see text for photon burst details).

The photon burst in Fig. 5A with 308 ms duration switched from initial P~0.68 (15 ms) to P~0.20 (63 ms), followed by a low intensity period of burst, then switched to P~0.18 (29 ms) and to P~0.86 (37 ms) before leaving the detection volume with P~0.27 (12 ms). The second example in Fig. 5B (221 ms) began with P~0.62 for 90 ms, switched to P~0.31 (20 ms) and then to P~0.81 (34 ms). In Fig. 5C, the 167 ms long photon bursts started at P~0.50 (28 ms), changed to P~0.24 (41 ms), then to P~0.59 (16 ms) and finally to P~0.22 (39 ms). The enzyme in Fig. 5D switched from P~0.75 (5 ms) to P~0.43 (7 ms) in a short 37 ms burst, and the enzyme in Fig. 5E changed proximity factors from P~0.42 (13 ms) to P~0.26 (46 ms) during a 138 ms observation time. Fig. 5F shows a photon burst with 197 ms duration, that apparently changed from P~0.17 to P~0.58 but with a maximum intensity of 170 counts per ms, indicating this was probably not a single FRET-labeled $F_oF_1$-ATP synthase.

To compare the conformational states and changes of the C-terminal part of ε in the holoenzyme, we plotted the manually-assigned FRET levels (proximity factors P) for both biochemical conditions in Fig. 6. In the absence of added nucleotides we found two broad maxima in the P histogram of 231 states in Fig. 6A. The low-FRET fraction with 0<P<0.3 will likely contain a few Atto488-only-labeled $F_oF_1$-ATP synthases (FRET donor only) including some enzymes that had the residual δ subunit labeled with Atto488. The second population (0.4<P<0.8) with maximum at P~0.65 could represent enzymes with ε in the 'up' conformation. In contrast, the proximity factor distribution in the presence of 1 mM MgATP (Fig. 6B) showed three distinguishable populations for the 128 states: one with FRET levels 0<P<0.3 as before, a second population with P values shifted to lower FRET efficiencies (0.35<P<0.6), and a third population with high FRET levels (0.7<P<0.9). The number of FRET levels in these two populations with P>0.3 seemed to be similar, but with a slightly higher occurrence of the lower FRET levels.

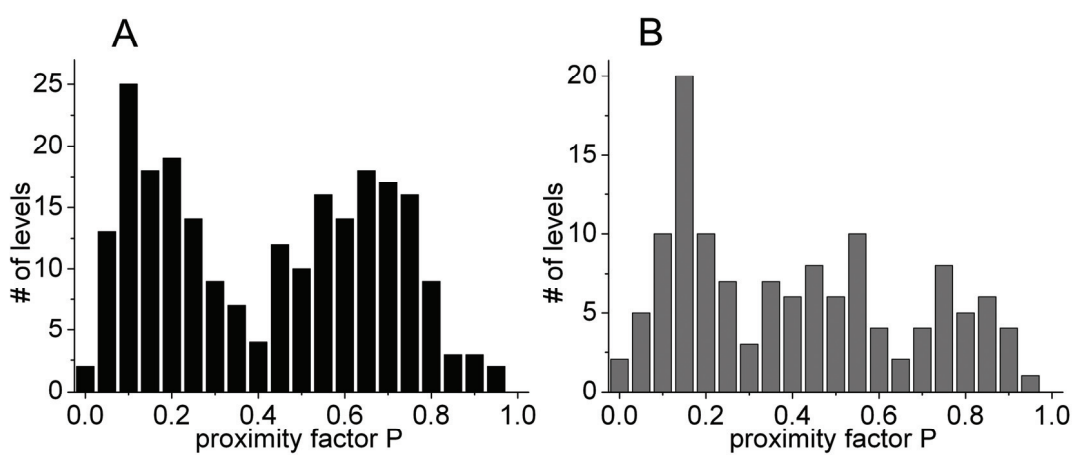

**Figure 6.** Histograms of proximity factors P for FRET-labeled reconstituted $F_oF_1$-ATP synthases in the absence of added nucleotides (**A**) and in the presence of 1 mM MgATP (**B**). The total numbers of manually assigned P levels were 231 in (**A**) for 1400 s recording time, and 128 in (**B**) for 650 s recording time.

## 4 DISCUSSION

Monitoring conformational changes within single reconstituted $F_oF_1$-ATP synthase required an *in vitro* assembly process using specifically labeled subunits and protein portions of this membrane enzyme from *E. coli*. Assembly procedures had been developed previously for single-molecule FRET measurements of subunit rotation in reconstituted $F_oF_1$-ATP synthase in buffer solution[18-23, 27, 28]. Binding constants for individual subunits as well as $F_1$ to $F_o$ assembly are in the 1-nM range or below[36, 64-68]. First we purified the $F_1$-γ108C portion for labeling with Atto488. This $F_1$ was depleted of δ and ε subunits. Separately we purified a cysteine mutant of the ε subunit, ε99C, for labeling with Atto647N. $F_1$-γ108-atto488 and ε99-atto647N were combined to yield $F_1$-γ108-atto488/ε99-atto647N as described previously[47] (see also S. D. Bockenhauer et al.[69], preprint available at http://arxiv.org/abs/1402.1845). To complete $F_1$ with all 9 subunits, we added an excess of additional δ that was overexpressed[44] and purified separately. The FRET-labeled $F_oF_1$-ATP synthase was obtained by rebinding $F_1$ to the non-labeled $F_o$ portion that was reconstituted in liposomes stochastically with less than one $F_o$ per vesicle. Successful assembly of the holoenzyme was proven by fluorescence correlation spectroscopy showing diffusion times for the FRET-labeled proteoliposomes in the range of 30 to 70 ms, i.e. ~100 times longer than a single rhodamine 110 molecule in buffer.

We analyzed the conformation of ε's C-terminal helices in $F_oF_1$-ATP synthase by confocal single-molecule FRET using a custom-designed basic microscope with two SPAD detectors[18, 61, 70]. However, to reveal the simultaneous attachment of both FRET donor and acceptor fluorophores on the individual enzyme, a more sophisticated laser excitation scheme with two alternating wavelengths is recommended. We have established such a confocal setup for duty cycle-optimized alternating laser excitation (DCO-ALEX[71-76]) using pulsed lasers but could not use this microscope for our measurements here. Therefore, low FRET efficiencies of photon bursts with proximity factors 0<P<0.2 might result from FRET donor-only labeled enzymes and should be omitted from further analysis. Very high FRET efficiencies in photon bursts as shown in Fig. 4A should be confirmed using the second pulsed laser in DCO-ALEX (at 635 nm) and fluorescence lifetime analysis of the FRET acceptor. Furthermore, single-molecule fluorescence anisotropies of FRET donor and acceptor in the local protein environments on γ and ε should be determined to ensure a reasonably high flexibility of the attached dyes to justify the assumption of $κ^2=2/3$ according to the Förster theory of FRET[77].

Single-molecule FRET measurements would benefit from longer observation times of individual enzymes. One elegant device to hold single proteoliposomes in solution is the anti-Brownian electrokinetic (ABEL) trap invented by A. E. Cohen and W. E. Moerner[78, 79]. Microfluidics confine diffusion to two dimensions, fluorescence is used to identify the position of the labeled molecule in real time, and a fast feedback[80] in μs pushes the object back to the center of confocal imaging using electric potentials generated with platinum electrodes. We have recently used the ABEL trap to analyze ε's CTD in single FRET-labeled $F_1$-γ108-atto488/ε99-atto647N in solution[69]. Constant fluorescence intensities of ABEL-trapped $F_1$ could be used to eliminate aggregates and other artifacts in the data set from further analysis. Due to long observation times of hundreds of ms (instead of only 2 to 5 ms for $F_1$ freely diffusing through the detection volume) we could identify FRET levels and fluctuations in the absence and presence of added MgATP and MgAMPPNP. This approach could also be used for the proteoliposomes measured here, and would improve the detection of fluctuating FRET levels. Furthermore, ambiguities due to manual assignment of FRET levels in the time traces and small smFRET data sets (sufficient for our proof-of-principle study here) have to be solved, for example, using automatic data analysis tools like Hidden Markov Models (HMM)[75, 81].

Finally, we want to improve the smFRET detection of ε's CTD conformational changes in single $F_oF_1$-ATP synthase by discrimination of rotating, active enzymes and ε-inhibited $F_oF_1$. Using the C-terminal cysteine mutant for specifically labeling subunit *a* in $F_o$ (as shown in Fig. 2) we will be able to simultaneously measure rotation of subunit γ and ε by smFRET to *a* as well as the conformation of ε's CTD by alternating lasers in nanoseconds for smFRET between γ108 and ε99. Similar to our first implementation of 3-color smFRET experiments[23] to monitor ε rotation and elastic twisting of the γ-ε-c rotor portion in single $F_oF_1$-ATP synthase[28, 82, 83], we aim at maximizing the obtainable information about intramolecular distance fluctuations within single enzymes one-at-a-time.

Despite the technical factors noted above, the preliminary smFRET results presented here indicate significant potential for further understanding the dynamic conformational changes of ε's CTD and how these correlate with function control of the intact bacterial ATP synthase. The results are especially interesting in comparison with our initial smFRET results with isolated $F_1$ that used the same labeled donor/acceptor pair, γ108C/ε99C[47]. First, the slower diffusion of $F_oF_1$-liposomes here allowed longer observation times and we were able to observe transitions between FRET states, especially upon addition of substrate MgATP; more extensive data sets should allow us to characterize the kinetics of ε's conformational changes under different functional conditions (*e.g.*, ATP synthesis *vs* hydrolysis). Second, without added nucleotide, soluble $F_1$/ε showed a FRET distribution predominated by one main peak at ~0.6 P, whereas $F_oF_1$-liposomes in this study showed a distinct, bimodal distribution between low- and high-FRET states (Fig. 6). This difference is expected from functional studies, since >90% of soluble $F_1$/ε complexes are in the ε-inhibited state[34, 36], whereas ATPase of $F_oF_1$ appears to be only ~50% inhibited by the ε CTD[35, 39]. Third, with MgATP added, soluble $F_1$/ε showed a bimodal FRET distribution but a trimodal distribution is observed here with $F_oF_1$-liposomes. This could indicate that association of $F_1$ with $F_o$ alters the kinetically significant intermediate states/positions of ε's CTD. This is not unexpected, since functional[36] and rotary-bead studies[37, 84] indicate that the ε CTD inserts into $F_1$'s central cavity at the catalytic dwell (+80° rotary angle), whereas the structure of ε-inhibited *E. coli* $F_1$ appears to be stopped after the next rotary sub-step (+120° rotary angle). Also, the transition from ε's 'down' state toward the 'up' state may involve an intermediate with a highly mobile CTD, which would likely put ε99C more distant from γ108C (on average), and could account for some of the lowest FRET levels observed here. Our future experiments with 3-color smFRET, including a probe on subunit *a* of $F_o$, should help further correlate these putative states of ε's CTD with functional and rotational states of the $F_oF_1$-ATP synthase.

**Acknowledgements**

We thank Marcus L. Hutcheon (now at Bristol-Myers Squibb, Syracuse, NY) for purification of *E. coli* $F_1$ proteins used in this study. We thank Professor Stanley D. Dunn (University of Western Ontario, London, Canada) for providing the plasmid pJC1. Financial support by NIG grant R01GM083088 to T. M. D. and by the Baden-Württemberg Stiftung (contract research project P-LS-Meth/6 in the program "Methods for Life Sciences") to M. B. is gratefully acknowledged. This work was supported in part by DFG grants BO 1891/15-1 and BO 1891/16-1 to M. B..

# 5 REFERENCES

[1] Weber, J., and Senior, A. E., "Catalytic mechanism of F1-ATPase," Biochim Biophys Acta 1319, 19-58 (1997).

[2] Boyer, P. D., "The ATP synthase--a splendid molecular machine," Annu Rev Biochem 66, 717-749 (1997).

[3] Duncan, T. M., Bulygin, V. V., Zhou, Y., Hutcheon, M. L., and Cross, R. L., "Rotation of subunits during catalysis by Escherichia coli F1-ATPase," Proc Natl Acad Sci U S A 92, 10964-10968 (1995).

[4] Zhou, Y., Duncan, T. M., Bulygin, V. V., Hutcheon, M. L., and Cross, R. L., "ATP hydrolysis by membrane-bound Escherichia coli F0F1 causes rotation of the gamma subunit relative to the beta subunits," Biochim Biophys Acta 1275, 96-100 (1996).

[5] Zhou, Y., Duncan, T. M., and Cross, R. L., "Subunit rotation in Escherichia coli FoF1-ATP synthase during oxidative phosphorylation," Proc Natl Acad Sci U S A 94, 10583-10587 (1997).

[6] Bulygin, V. V., Duncan, T. M., and Cross, R. L., "Rotation of the epsilon subunit during catalysis by Escherichia coli FOF1-ATP synthase," J Biol Chem 273, 31765-31769 (1998).

[7] Hutcheon, M. L., Duncan, T. M., Ngai, H., and Cross, R. L., "Energy-driven subunit rotation at the interface between subunit a and the c oligomer in the F(O) sector of Escherichia coli ATP synthase," Proc Natl Acad Sci U S A 98, 8519-8524 (2001).

[8] Sabbert, D., Engelbrecht, S., and Junge, W., "Intersubunit rotation in active F-ATPase," Nature 381, 623-625 (1996).

[9] Noji, H., Yasuda, R., Yoshida, M., and Kinosita, K., Jr., "Direct observation of the rotation of F1-ATPase," Nature 386, 299-302 (1997).

[10] Yasuda, R., Noji, H., Yoshida, M., Kinosita, K., Jr., and Itoh, H., "Resolution of distinct rotational substeps by submillisecond kinetic analysis of F1-ATPase," Nature 410, 898-904 (2001).

[11] Bilyard, T., Nakanishi-Matsui, M., Steel, B. C., Pilizota, T., Nord, A. L., Hosokawa, H., Futai, M., and Berry, R. M., "High-resolution single-molecule characterization of the enzymatic states in Escherichia coli F1-ATPase," Philosophical Transactions of the Royal Society B: Biological Sciences 368, 20120023 (2013).

[12] Spetzler, D., York, J., Daniel, D., Fromme, R., Lowry, D., and Frasch, W., "Microsecond time scale rotation measurements of single F1-ATPase molecules," Biochemistry 45, 3117-3124 (2006).

[13] Ishmukhametov, R., Hornung, T., Spetzler, D., and Frasch, W. D., "Direct observation of stepped proteolipid ring rotation in E. coli FF-ATP synthase," Embo J 29, 3911-3923 (2010).

[14] Watanabe, R., Tabata, K. V., Iino, R., Ueno, H., Iwamoto, M., Oiki, S., and Noji, H., "Biased Brownian stepping rotation of FoF1-ATP synthase driven by proton motive force," Nat Commun 4, 1631 (2013).

[15] Adachi, K., Yasuda, R., Noji, H., Itoh, H., Harada, Y., Yoshida, M., and Kinosita, K., Jr., "Stepping rotation of F1-ATPase visualized through angle-resolved single-fluorophore imaging," Proc Natl Acad Sci U S A 97, 7243-7247 (2000).

[16] Hasler, K., Engelbrecht, S., and Junge, W., "Three-stepped rotation of subunits gamma and epsilon in single molecules of F-ATPase as revealed by polarized, confocal fluorometry," FEBS Lett 426, 301-304 (1998).

[17] Kaim, G., Prummer, M., Sick, B., Zumofen, G., Renn, A., Wild, U. P., and Dimroth, P., "Coupled rotation within single F0F1 enzyme complexes during ATP synthesis or hydrolysis," FEBS Lett 525, 156-163 (2002).

[18] Borsch, M., Diez, M., Zimmermann, B., Reuter, R., and Graber, P., "Stepwise rotation of the gamma-subunit of EF(0)F(1)-ATP synthase observed by intramolecular single-molecule fluorescence resonance energy transfer," FEBS Lett 527, 147-152 (2002).

[19] Borsch, M., Diez, M., Zimmermann, B., Trost, M., Steigmiller, S., and Graber, P., "Stepwise rotation of the gamma-subunit of EFoF1-ATP synthase during ATP synthesis: a single-molecule FRET approach," Proc. SPIE 4962, 11-21 (2003).

[20] Diez, M., Zimmermann, B., Borsch, M., Konig, M., Schweinberger, E., Steigmiller, S., Reuter, R., Felekyan, S., Kudryavtsev, V., Seidel, C. A., and Graber, P., "Proton-powered subunit rotation in single membrane-bound FoF1-ATP synthase," Nat Struct Mol Biol 11, 135-141 (2004).

[21] Zimmermann, B., Diez, M., Zarrabi, N., Graber, P., and Borsch, M., "Movements of the epsilon-subunit during catalysis and activation in single membrane-bound H(+)-ATP synthase," Embo J 24, 2053-2063 (2005).

[22] Duser, M. G., Zarrabi, N., Cipriano, D. J., Ernst, S., Glick, G. D., Dunn, S. D., and Borsch, M., "36 degrees step size of proton-driven c-ring rotation in FoF1-ATP synthase," Embo J 28, 2689-2696 (2009).

[23] Ernst, S., Duser, M. G., Zarrabi, N., and Borsch, M., "Three-color Förster resonance energy transfer within single FoF1-ATP synthases: monitoring elastic deformations of the rotary double motor in real time," J Biomed Opt 17, 011004 (2012).

[24] Yasuda, R., Masaike, T., Adachi, K., Noji, H., Itoh, H., and Kinosita, K., Jr., "The ATP-waiting conformation of rotating F1-ATPase revealed by single-pair fluorescence resonance energy transfer," Proc Natl Acad Sci U S A 100, 9314-9318 (2003).

[25] Zarrabi, N., Zimmermann, B., Diez, M., Graber, P., Wrachtrup, J., and Borsch, M., "Asymmetry of rotational catalysis of single membrane-bound F0F1-ATP synthase," Proc. SPIE 5699, 175-188 (2005).

[26] Johnson, K. M., Swenson, L., Opipari, A. W., Jr., Reuter, R., Zarrabi, N., Fierke, C. A., Borsch, M., and Glick, G. D., "Mechanistic basis for differential inhibition of the F(1)F(o)-ATPase by aurovertin," Biopolymers 91, 830-840 (2009).

[27] Duser, M. G., Bi, Y., Zarrabi, N., Dunn, S. D., and Borsch, M., "The proton-translocating a subunit of F0F1-ATP synthase is allocated asymmetrically to the peripheral stalk," J Biol Chem 283, 33602-33610 (2008).

[28] Ernst, S., Duser, M. G., Zarrabi, N., Dunn, S. D., and Borsch, M., "Elastic deformations of the rotary double motor of single FoF1-ATP synthases detected in real time by Förster resonance energy transfer," Biochimica et Biophysica Acta (BBA) - Bioenergetics 1817, 1722-1731 (2012).

[29] Cingolani, G., and Duncan, T. M., "Structure of the ATP synthase catalytic complex (F(1)) from Escherichia coli in an autoinhibited conformation," Nat Struct Mol Biol 18, 701-707 (2011).

[30] Roy, A., Hutcheon, M. L., Duncan, T. M., and Cingolani, G., "Improved crystallization of Escherichia coli ATP synthase catalytic complex (F1) by introducing a phosphomimetic mutation in subunit epsilon," Acta Crystallogr Sect F Struct Biol Cryst Commun 68, 1229-1233 (2012).

[31] Wachter, A., Bi, Y., Dunn, S. D., Cain, B. D., Sielaff, H., Wintermann, F., Engelbrecht, S., and Junge, W., "Two rotary motors in F-ATP synthase are elastically coupled by a flexible rotor and a stiff stator stalk," Proc Natl Acad Sci U S A 108, 3924-3929 (2011).

[32] Bottcher, B., Bertsche, I., Reuter, R., and Graber, P., "Direct visualisation of conformational changes in EF(0)F(1) by electron microscopy," J Mol Biol 296, 449-457 (2000).

[33] Wilkens, S., and Capaldi, R. A., "Solution structure of the epsilon subunit of the F1-ATPase from Escherichia coli and interactions of this subunit with beta subunits in the complex," J Biol Chem 273, 26645-26651 (1998).

[34] Nakanishi-Matsui, M., Kashiwagi, S., Hosokawa, H., Cipriano, D. J., Dunn, S. D., Wada, Y., and Futai, M., "Stochastic high-speed rotation of Escherichia coli ATP synthase F1 sector: the epsilon subunit-sensitive rotation," J Biol Chem 281, 4126-4131 (2006).

[35] Iino, R., Hasegawa, R., Tabata, K. V., and Noji, H., "Mechanism of inhibition by C-terminal alpha-helices of the epsilon subunit of Escherichia coli FoF1-ATP synthase," J Biol Chem 284, 17457-17464 (2009).

[36] Shah, N. B., Hutcheon, M. L., Haarer, B. K., and Duncan, T. M., "F1-ATPase of Escherichia coli: the epsilon- inhibited state forms after ATP hydrolysis, is distinct from the ADP-inhibited state, and responds dynamically to catalytic site ligands," J Biol Chem 288, 9383-9395 (2013).

[37] Sekiya, M., Hosokawa, H., Nakanishi-Matsui, M., Al-Shawi, M. K., Nakamoto, R. K., and Futai, M., "Single molecule behavior of inhibited and active states of Escherichia coli ATP synthase F1 rotation," J Biol Chem 285, 42058-42067 (2010).

[38] Uhlin, U., Cox, G. B., and Guss, J. M., "Crystal structure of the epsilon subunit of the proton-translocating ATP synthase from Escherichia coli," Structure 5, 1219-1230 (1997).

[39] Schulenberg, B., and Capaldi, R. A., "The epsilon subunit of the F(1)F(0) complex of Escherichia coli. cross-linking studies show the same structure in situ as when isolated," J Biol Chem 274, 28351-28355 (1999).

[40] Wise, J. G., "Site-directed mutagenesis of the conserved beta subunit tyrosine 331 of Escherichia coli ATP synthase yields catalytically active enzymes," J Biol Chem 265, 10403-10409 (1990).

[41] Peneva, K., Mihov, G., Herrmann, A., Zarrabi, N., Borsch, M., Duncan, T. M., and Mullen, K., "Exploiting the Nitrilotriacetic Acid Moiety for Biolabeling with Ultrastable Perylene Dyes," J Am Chem Soc 130, 5398-5399 (2008).

[42] Kuo, P. H., Ketchum, C. J., and Nakamoto, R. K., "Stability and functionality of cysteine-less F(0)F1 ATP synthase from Escherichia coli," FEBS Lett 426, 217-220 (1998).

[43] Andrews, S. H., Peskova, Y. B., Polar, M. K., Herlihy, V. B., and Nakamoto, R. K., "Conformation of the gamma subunit at the gamma-epsilon-c interface in the complete Escherichia coli F(1)-ATPase complex by site-directed spin labeling," Biochemistry 40, 10664-10670 (2001).

[44] Dunn, S. D., and Chandler, J., "Characterization of a b2delta complex from Escherichia coli ATP synthase," J Biol Chem 273, 8646-8651 (1998).

[45] Schaefer, E. M., Hartz, D., Gold, L., and Simoni, R. D., "Ribosome-binding sites and RNA-processing sites in the transcript of the Escherichia coli unc operon," Journal of Bacteriology 171, 3901-3908 (1989).

[46] Dunn, S. D., "Removal of the epsilon subunit from Escherichia coli F1-ATPase using monoclonal anti-epsilon antibody affinity chromatography," Anal Biochem 159, 35-42 (1986).

[47] Borsch, M., and Duncan, T. M., "Spotlighting motors and controls of single FoF1-ATP synthase," Biochem Soc Trans 41, 1219-1226 (2013).

[48] Borsch, M., Turina, P., Eggeling, C., Fries, J. R., Seidel, C. A., Labahn, A., and Graber, P., "Conformational changes of the H+-ATPase from Escherichia coli upon nucleotide binding detected by single molecule fluorescence," FEBS Lett 437, 251-254 (1998).

[49] Aggeler, R., Chicas-Cruz, K., Cai, S. X., Keana, J. F. W., and Capaldi, R. A., "Introduction of reactive cysteine residues in the .epsilon. subunit of Escherichia coli F1 ATPase, modification of these sites with (azidotetrafluorophenyl)maleimides, and examination of changes in the binding of the .epsilon. subunit when different nucleotides are in catalytic sites," Biochemistry 31, 2956-2961 (1992).

[50] Aggeler, R., Ogilvie, I., and Capaldi, R. A., "Rotation of a gamma-epsilon subunit domain in the Escherichia coli F1F0-ATP synthase complex. The gamma-epsilon subunits are essentially randomly distributed relative to the alpha3beta3delta domain in the intact complex," J Biol Chem 272, 19621-19624 (1997).

[51] Friedl, P., and Schairer, H. U., "Preparation and reconstitution of F1F0 and F0 from Escherichia coli," Methods Enzymol 126, 579-588 (1986).

[52] Heitkamp, T., Sielaff, H., Korn, A., Renz, M., Zarrabi, N., and Borsch, M., "Monitoring subunit rotation in single FRET-labeled FoF1-ATP synthase in an anti-Brownian electrokinetic trap," Proc. SPIE 8588, 85880Q (2013).

[53] Fischer, S., Etzold, C., Turina, P., Deckers-Hebestreit, G., Altendorf, K., and Graber, P., "ATP synthesis catalyzed by the ATP synthase of Escherichia coli reconstituted into liposomes," Eur J Biochem 225, 167-172 (1994).

[54] Fischer, S., and Graber, P., "Comparison of DeltapH- and Delta***φ***-driven ATP synthesis catalyzed by the H(+)-ATPases from Escherichia coli or chloroplasts reconstituted into liposomes," FEBS Lett 457, 327-332 (1999).

[55] Holloway, P. W., "A simple procedure for removal of Triton X-100 from protein samples," Anal Biochem 53, 304-308 (1973).

[56] Lotscher, H. R., deJong, C., and Capaldi, R. A., "Modification of the F0 portion of the H+-translocating adenosinetriphosphatase complex of Escherichia coli by the water-soluble carbodiimide 1-ethyl-3-[3-(dimethylamino)propyl]carbodiimide and effect on the proton channeling function," Biochemistry 23, 4128-4134 (1984).

[57] Popov, N., Schmitt, M., Schulzeck, S., and Matthies, H., "Eine störungsfreie Mikromethode zur Bestimmung des Proteingehaltes in Gewebehomogenaten," Acta Biol Med Ger 34, 1441-1446 (1975).

[58] Schagger, H., and von Jagow, G., "Tricine-sodium dodecyl sulfate-polyacrylamide gel electrophoresis for the separation of proteins in the range from 1 to 100 kDa," Anal Biochem 166, 368-379 (1987).

[59] Heukeshoven, J., and Dernick, R., "Simplified Method for Silver Staining of Proteins in Polyacrylamide Gels and the Mechanism of Silver Staining," Electrophoresis 6, 103-112 (1985).

[60] Weber, K., and Osborn, M., "The reliability of molecular weight determinations by dodecyl sulfate-polyacrylamide gel electrophoresis," J Biol Chem 244, 4406-4412 (1969).

[61] Zarrabi, N., Clausen, C., Duser, M. G., and Borsch, M., "Manipulating freely diffusing single 20-nm particles in an Anti-Brownian Electrokinetic Trap (ABELtrap)," Proc. SPIE 8587, 85870L (2013).

[62] Sielaff, H., Heitkamp, T., Zappe, A., Zarrabi, N., and Borsch, M., "Subunit rotation in single FRET-labeled F1-ATPase hold in solution by an anti-Brownian electrokinetic trap," Proc. SPIE 8590, 859008 (2013).

[63] Laemmli, U. K., "Cleavage of structural proteins during the assembly of the head of bacteriophage T4," Nature 227, 680-685 (1970).

[64] Diez, M., Borsch, M., Zimmermann, B., Turina, P., Dunn, S. D., and Graber, P., "Binding of the b-subunit in the ATP synthase from Escherichia coli," Biochemistry 43, 1054-1064 (2004).

[65] Krebstakies, T., Zimmermann, B., Graber, P., Altendorf, K., Borsch, M., and Greie, J. C., "Both rotor and stator subunits are necessary for efficient binding of F1 to F0 in functionally assembled Escherichia coli ATP synthase," J Biol Chem 280, 33338-33345 (2005).

[66] Hasler, K., Panke, O., and Junge, W., "On the stator of rotary ATP synthase: the binding strength of subunit delta to (alpha beta)3 as determined by fluorescence correlation spectroscopy," Biochemistry 38, 13759-13765 (1999).

[67] Weber, J., Wilke-Mounts, S., Nadanaciva, S., and Senior, A. E., "Quantitative determination of direct binding of b subunit to F1 in Escherichia coli F1F0-ATP synthase," J Biol Chem 279, 11253-11258 (2004).

[68] Weber, J., Wilke-Mounts, S., and Senior, A. E., "Quantitative determination of binding affinity of delta-subunit in Escherichia coli F1-ATPase: effects of mutation, Mg2+, and pH on Kd," J Biol Chem 277, 18390-18396 (2002).

[69] Bockenhauer, S. D., Duncan, T. M., Moerner, W. E., and Borsch, M., "The regulatory switch of F1-ATPase studied by single-molecule FRET in the ABEL Trap," Proc. SPIE 8950, in press (2014).

[70] Steigmiller, S., Zimmermann, B., Diez, M., Borsch, M., and Graber, P., "Binding of single nucleotides to H+-ATP synthases observed by fluorescence resonance energy transfer," Bioelectrochemistry 63, 79-85 (2004).

[71] Zarrabi, N., Duser, M. G., Ernst, S., Reuter, R., Glick, G. D., Dunn, S. D., Wrachtrup, J., and Borsch, M., "Monitoring the rotary motors of single FoF1-ATP synthase by synchronized multi channel TCSPC," Proc. SPIE 6771, 67710F (2007).

[72] Zarrabi, N., Ernst, S., Duser, M. G., Golovina-Leiker, A., Becker, W., Erdmann, R., Dunn, S. D., and Borsch, M., "Simultaneous monitoring of the two coupled motors of a single FoF1-ATP synthase by three-color FRET using duty cycle-optimized triple-ALEX," Proc. SPIE 7185, 718505 (2009).

[73] Verhalen, B., Ernst, S., Borsch, M., and Wilkens, S., "Dynamic ligand induced conformational rearrangements in P-glycoprotein as probed by fluorescence resonance energy transfer spectroscopy," Journal of Biological Chemistry 287, 1112-1127 (2012).

[74]   Winterfeld, S., Ernst, S., Borsch, M., Gerken, U., and Kuhn, A., "Real time observation of single membrane protein insertion events by the Escherichia coli insertase YidC," PLoS One 8, e59023 (2013).
[75]   Zarrabi, N., Ernst, S., Verhalen, B., Wilkens, S., and Borsch, M., "Analyzing conformational dynamics of single P-glycoprotein transporters by Förster resonance energy transfer using hidden Markov models," Methods 65, in press (2014).
[76]   Zarrabi, N., Heitkamp, T., Greie, J.-C., and Borsch, M., "Monitoring the conformational dynamics of a single potassium transporter by ALEX-FRET," Proc. SPIE 6862, 68620M (2008).
[77]   Förster, T., "Energiewanderung Und Fluoreszenz," Naturwissenschaften 33, 166-175 (1946).
[78]   Cohen, A. E., and Moerner, W. E., "Method for trapping and manipulating nanoscale objects in solution," Appl. Phys. Lett. 86, 093109 (2005).
[79]   Cohen, A. E., and Moerner, W. E., "Controlling Brownian motion of single protein molecules and single fluorophores in aqueous buffer," Opt Express 16, 6941-6956 (2008).
[80]   Wang, Q., and Moerner, W. E., "Optimal strategy for trapping single fluorescent molecules in solution using the ABEL trap," Appl Phys B 99, 23-30 (2010).
[81]   Zarrabi, N., Duser, M. G., Reuter, R., Dunn, S. D., Wrachtrup, J., and Borsch, M., "Detecting substeps in the rotary motors of FoF1-ATP synthase by Hidden Markov Models," Proc. SPIE 6444, 64440E (2007).
[82]   Sielaff, H., and Borsch, M., "Twisting and subunit rotation in single FOF1-ATP synthase," Phil Trans R Soc B 368, 20120024 (2013).
[83]   Borsch, M., "Microscopy of single FoF1-ATP synthases— The unraveling of motors, gears, and controls," IUBMB Life 65, 227-237 (2013).
[84]   Tsumuraya, M., Furuike, S., Adachi, K., Kinosita, K., Jr., and Yoshida, M., "Effect of epsilon subunit on the rotation of thermophilic Bacillus F1-ATPase," FEBS Lett 583, 1121-1126 (2009).